\documentclass[a4paper,11pt]{article}
\usepackage{geometry}
\usepackage[T1]{fontenc}
\usepackage[utf8]{inputenc}
\usepackage[english]{babel}
\usepackage{lmodern}
\usepackage{graphicx,graphics}
\usepackage{color}
\usepackage{url} 
\usepackage{epsf,epsfig}
\usepackage{subcaption}
\usepackage{amsmath,amsfonts,amssymb,slashed}
\usepackage{bbold}
\usepackage[colorlinks=true, urlcolor=blue]{hyperref}
\everymath{\displaystyle}
\usepackage{authblk}

\title{\textbf{Out-of-equilibrium chiral condensate in AdS/QCD}}
\author[1]{Floriana Giannuzzi\thanks{\href{{mailto:floriana.giannuzzi@ba.infn.it}}{floriana.giannuzzi@ba.infn.it}}}
\author[1]{Stefano Nicotri\thanks{\href{mailto:nicotri@infn.it}{nicotri@infn.it}}}
\affil[1]{\small \emph{INFN -- Istituto Nazionale di Fisica Nucleare -- Sezione di Bari} \protect\\ \emph{Via Orabona 4, 70125, Bari, Italy}}
\date{}

\begin{document}
\begin{flushright}{BARI-TH/771-25} \end{flushright}
{\let\newpage\relax\maketitle}

\maketitle

\begin{abstract}
We study the chiral condensate at finite temperature in AdS/QCD in a time-dependent
background, in which the position of the black-hole horizon $z_h$ changes with time, producing an increasing or decreasing temperature. 
Conformal invariance is broken, as in the soft-wall model, by a static quadratic dilaton. 
Two different scenarios are analysed: in the first a general power-law time dependence is assumed for $z_h$, while in the second the energy-momentum tensor at late times reproduces the one found in viscous hydrodynamics. 
Depending on the rate at which the system evolves, the transition shifts toward lower temperatures.
If the chiral condensate is far from equilibrium at low temperatures, oscillations around its equilibrium value are observed before thermalization.
A prethermalization stage is found in the chiral limit if the initial condition is set at a temperature close to the critical one.
In the hydrodynamic setup the evolution of the medium is slow enough, and the chiral condensate soon matches the equilibrium curve.
\end{abstract}

\section{Introduction}
Quantum Chromodynamics (QCD) has a rich phase diagram in the chemical potential -- temperature plane, each phase characterised by different degrees of freedom. 
In the hadronic phase, realised at small temperature and chemical potential, quarks are confined in colorless states, the hadrons, and chiral symmetry is broken. 
At higher values of temperature and chemical potential QCD is in the deconfined quark-gluon plasma (QGP) phase  in which quarks and gluons are free and chiral symmetry is restored.    
For small values of the chemical potential a crossover separates the two phases, becoming a second-order phase transition in the case of two massless quarks.
Heavy-ion collision experiments at the LHC and RHIC have provided a wealth of data to investigate the fundamental properties of QGP by studying the dynamical evolution of the fireballs generated in the collisions. 
Data are well described by viscous hydrodynamics, showing that QGP behaves as a nearly perfect fluid with very low viscosity, and has collective flow patterns which indicate strong interactions among quarks and gluons. 
These flow patterns are characterized by anisotropic flow coefficients, which provide information about the initial geometry and the subsequent evolution of the QGP.

Many  theoretical models developed to study QGP involve a static system at thermal equilibrium, with a  fixed temperature.
Anyway, the strongly interacting matter produced in heavy-ion collisions undergoes rapid expansion and cooling, so out-of-equilibrium effects should be taken into account to provide a realistic theoretical description of the phenomenon, in particular when fast-changing conditions are involved, thus making the equilibrium approximation inappropriate.
In this paper we use AdS/QCD to dynamically study the chiral phase transition at zero chemical potential, with out-of-equilibrium effects  included by assuming a time-dependent metric in the bulk and either on-shell or perturbed initial conditions. 

In the past, some theoretical efforts have already been devoted to the  analysis of the evolution of the chiral condensate and the chiral transition in out-of-equilibrium dynamical systems in some phenomenological models \cite{Meistrenko:2020nwx,Rajagopal:2019xwg,Carlomagno:2018ogx,Grosse:2007ty,Wesp:2017tze,Rajagopal:1995bc,Abada:1994mf}.
In \cite{Bedaque:1993fa} the dynamics of a second order chiral phase transition is studied in the Nambu-Jona-Lasinio model, and in \cite{Wang:2020wwm} the chiral phase transition in an expanding quark-antiquark system is investigated solving the Vlasov and gap equations.  
Chiral transition and the chiral condensate have been investigated using a linear $\sigma$ model in \cite{Mishustin:1997iz}, observing strong oscillations of the chiral field during the transition.

Some approaches involving holographic methods to dynamically study the chiral condensate have been developed as well. 
In \cite{Evans:2010xs}  the  chiral phase transition is studied by placing probe D7 branes in a time-dependent geometry describing a growing/shrinking black hole in AdS with a background magnetic field, which is needed to induce chiral symmetry breaking in this framework.
In \cite{Cao:2022mep,Zheng:2024rzl} the holographic soft-wall model has been used to study the evolution of the chiral condensate by considering a time-dependent scalar field in a static black-hole geometry, perturbing the intial conditions.
In this paper we adopt a bottom-up holographic approach and consider different time-dependent geometries, in which the black-hole horizon moves either following a generic power-law behaviour or according to viscous hydrodynamics, thus changing the effective temperature of the system. A static dilaton field is introduced to break conformal invariance, and the dynamics of the scalar field containing the chiral condensate is studied.

The paper is organized as follows: in Sec.~\ref{sec:model} we introduce the holographic model we use to study a time-dependent chiral condensate; in Sec.~\ref{sec:static_case} we briefly review the results of the model in the static case (at equilibrium), in which the metric is time-independent; in Sec.~\ref{sec:dynamic_case} we introduce two time-dependent metrics we use to study the dynamics of the chiral condensate, one in which the temperature follows a power-law variation with time, described in Sec.~\ref{sec:power_law}, the other describing a hydrodinamic-like regime, investigated in Sec.~\ref{sec:hydro}. In Sec.~\ref{sec:conclusions} we summarize our conclusions and outlook.

\section{The model}\label{sec:model}
In order to study a dynamical chiral phase transition, we consider a time-dependent AdS-like geometry.
The starting point is a general five-dimensional metric in Eddington-Finkelstein coordinates:
\begin{equation}
    ds^{2}=-A(z,v)dv^{2}+\Sigma(z,v)^{2}e^{B(z,v)}d {x}_{\perp}^{2}+\Sigma(z,v)^{2}e^{-2B(z,v)}dy^{2}-\frac{2}{z^2}dv dz \,\,\,,
    \label{metric5D}
\end{equation}
respecting spatial $3d$ translation invariance, and $O(2)$ rotation invariance in the plane $x_\perp=\{x^1,x^2\}$ \cite{Chesler:2008hg}.
For a purely AdS metric the metric functions are: $A(z,v)=1/z^2$, $\Sigma(z,v)=1/z$ and $B(z,v)=0$. In an AdS space with a black hole the metric has: $A(z)=1/z^2 (1-z^4/z_h^4)$, where $z_h$ is the position of the black-hole horizon, related to the temperature $T$ of the dual gauge theory by $T=1/(\pi z_h)$.

In the holographic model considered in this work, chiral $U(n_f)_L\times U(n_f)_R$ symmetry of QCD, where $n_f$ is the number of quark flavors, is implemented via a dual local (gauge) symmetry in the bulk, with gauge fields $A_L$ and $A_R$.  
The scalar field  dual to the $q\bar q$ operator is $X=e^{i\pi} (X_0+S) e^{i\pi}$, where  the field $\pi(z,v,\vec x)$ describes pseudoscalar mesons and $S(z,v,\vec x)$ scalar mesons \cite{Colangelo:2008us,Giannuzzi:2021euy}. 
The vacuum expectation value (vev) $X_0(z,v)$ contains the scalar condensate, responsible for the breaking of such symmetry in a Higgs-like fashion.
For $n_f=2$ identical quarks, the vev is given by $X_0(z,v)\propto X_q(z,v) \mbox{diag}(1,1)$.
The breaking of conformal symmetry is implemented introducing a background dilaton field $\phi(z)=c^2 z^2$ which generates the QCD mass scale $\Lambda_{\mathrm{QCD}}$ \cite{Karch:2006pv}.
We work in the probe approximation and neglect the backreaction of the flavor fields and dilaton to the metric.
In the standard soft-wall model the chiral condensate is proportional to the quark mass, so it vanishes in the chiral limit $m_q\to0$. 
Following Ref. \cite{Cao:2020ryx}, in order to guarantee the existence of a finite chiral condensate at zero temperature also in the case of vanishing quark mass, a quartic term in the potential and an anomalous dimension (depending on the holographic coordinate $z$) in the five-dimensional mass of the $X$ field
\begin{equation}
    m_5^2=-3-\mu^2 z^2
\end{equation}
are introduced, resulting in the full action for $X_q(z,v)$ \cite{Erlich:2005qh,Colangelo:2011sr}:
\begin{equation}
    S \propto \int d^5x\, \sqrt{-g}\, e^{-\phi(z)} \,  \left[ (\partial_M X_q)(\partial^M X_q)  + m_5^2 X_q^2 +\lambda X_q^4  \right]\,,
\end{equation}
and equation of motion:
\begin{equation}\label{eq:eomgen}
    \partial^M(\sqrt{-g}\, e^{-\phi(z)} \, \partial_M X_q(z,v)) - m_5^2 \sqrt{-g}\, e^{-\phi(z)} \,X_q(z,v) - 2\lambda \sqrt{-g}\, e^{-\phi(z)} \, X_q(z,v)^3=0 \,.
\end{equation}
The values of the parameters we use are: $c=440$ MeV, $\mu=1450$ MeV, $\lambda=20$ \cite{Cao:2020ryx}.
In the following, we will briefly describe the static (equilibrium) case first, and then focus on the dynamic case, in which the position of the black-hole horizon changes with time, effectively taking the system out of equilibrium.
We will study the behavior of the chiral condensate in both cases.

\section{Chiral condensate in the static case}\label{sec:static_case}
\begin{figure}
    \begin{center}
    \includegraphics[width=9cm]{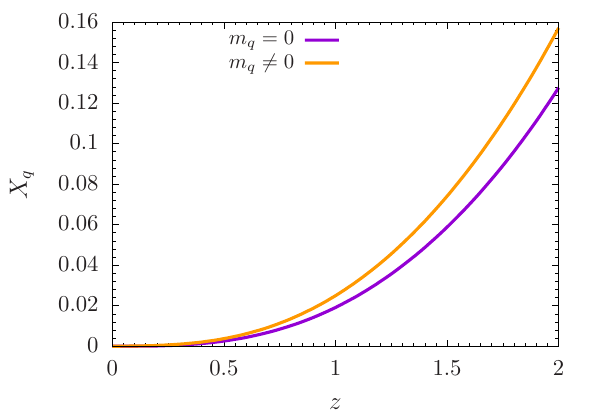}
    \end{center}
    \caption{Profile of  $X_q(z)$ for $z_h=2$ GeV$^{-1}$, and for $m_q=0$ (purple curve) and $m_q=3.22$ MeV (orange curve).}
    \label{fig:Xini}
\end{figure}

\begin{figure}[h!]
    \begin{center}
    \includegraphics[width=9cm]{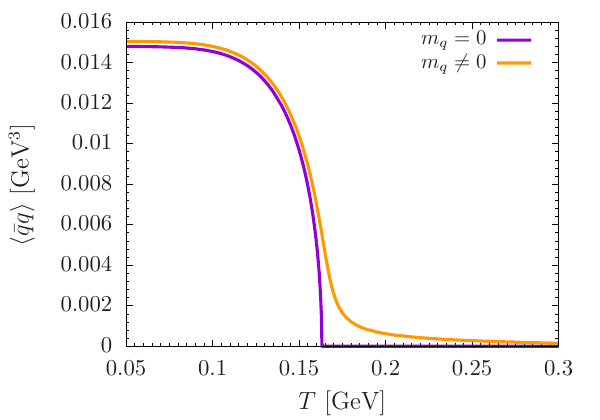}
    \end{center}
    \caption{Chiral condensate vs temperature in the static case for $m_q=0$ (purple curve) and $m_q=3.22$ MeV (orange curve).}
    \label{fig:sigmastatic}
\end{figure}

In the metric with a black hole, having $A(z)=1/z^2 (1-z^4/z_h^4)$, $\Sigma(z,v)=1/z$ and $B(z,v)=0$, the time-independent equation of motion for the scalar field $X_q(z,v)=X_q(z)$ is:
\begin{equation}\label{eq:eomstatic}
    \left(1-\frac{z^4}{z_h^4}\right) X_q''(z) +\left( -\frac{3}{z} - \frac{ z^3}{z_h^4} - 2 c^2 z \left( 1-\frac{z^4}{z_h^4} \right) \right)  X_q'(z) -\frac{1}{z^2} m_5^2 X_q(z) -\frac{2}{z^2}\lambda X_q(z)^3=0\,.
\end{equation}
It can be solved using a shooting method imposing $X_q(0)=0$ and requiring the solution to have a regular behaviour as $z\to z_h$ \cite{Colangelo:2011sr,Cao:2022mep}. 
In Fig. \ref{fig:Xini} the solutions at $z_h=2$ GeV$^{-1}$ for $m_q=0$ and $m_q\neq 0$ are shown.
The behavior of the field near the boundary $z \to 0$ is \cite{Cao:2022mep}
\begin{equation}\label{eq:lowz}
    X_q(z) \xrightarrow[z\to 0]{} m_q\,\gamma\, z+\sigma\, z^3  +\left( c^2 m_q\gamma + (m_q\gamma)^3 \lambda - \frac{1}{2} m_q\gamma \mu_c^2\right) z^3 \log z+ \mathcal{O}(z^5)
\end{equation}
where $\gamma=\sqrt{N_c}/2\pi$ and $\sigma$ is related to the chiral condensate at equilibrium: $\langle \bar q q\rangle= \gamma\sigma$ \cite{Giannuzzi:2021euy}. 
By fitting the numerical solution to \eqref{eq:lowz} for $z\to 0$ the value of $\sigma$ can be obtained for any fixed temperature. Such values are shown  in Fig. \ref{fig:sigmastatic} for $m_q=0$ and $m_q=3.22$ MeV at different temperatures $T$, reproducing the result of Ref. \cite{Cao:2020ryx}. 
In the chiral limit a second-order transition from $\sigma\neq 0$ at $T<T_c$ to $\sigma= 0$ at $T>T_c$ is observed, with $T_c\sim 163$ MeV. 
For $m_q>0$ a crossover is obtained instead.
In the chiral limit the value of $\sigma$ at zero temperature is $\sigma_0=0.0536$ GeV$^3$, corresponding to a chiral condensate $\langle \bar q q\rangle= (245 \mbox{ MeV})^3$.
For $m_q=3.22$ MeV, the value of $\sigma$ at zero temperature is $\sigma_0=0.0545$ GeV$^3$, corresponding to a chiral condensate $\langle \bar q q\rangle= (247 \mbox{ MeV})^3$.

\section{Time-dependent chiral condensate}\label{sec:dynamic_case}
Let us consider an out-of-equilibrium system in which the position of the black-hole horizon changes with time ($z_h(v)$). 
We change variable and define the new set $(u,v')$ with 
\begin{equation}
    \begin{cases}
        u=\frac{z}{z_h(v)} &\\
        v^\prime = v
    \end{cases}
\end{equation}
such that the horizon is at $u=1$ and $u \in [0,1]$.
For clarity of notation, we will continue calling the new $v^\prime$ variable $v$.
Using the metric in Eq. \eqref{metric5D}, the equation of motion \eqref{eq:eomgen} for the scalar field becomes:
\begin{eqnarray}
    && X_q^{\prime\prime} \left(2 u \dot{z}_h+A z_h^2 u^2 \right) + z_h \dot{X}_q \left(-3 \frac{\Sigma^\prime}{\Sigma}+\phi^\prime \right) -2 z_h \dot{X}_q^\prime + \nonumber\\
    && + X_q^\prime \left( 6 \dot{z}_h \frac{\Sigma^\prime}{\Sigma} u- u \dot{z}_h \phi^\prime +2 \dot{z}_h+3 u^2 z_h^2 A \frac{\Sigma^\prime}{\Sigma}-u^2 z_h^2 A \phi^\prime +u^2 z_h^2 A^\prime + 2 u z_h^2 A -3 z_h \frac{\dot{\Sigma}}{\Sigma} \right) +\nonumber\\
    &&  - \frac{m_5^2}{u^2}  X_q-\frac{2 \lambda  }{u^2} X_q^3= 0 \,,
\end{eqnarray}
where the prime indicates a derivative w.r.t. $u$ and the dot w.r.t. $v$. 
We solve this partial differential equation using the Chebyshev pseudospectral method \cite{trefethen2000spectral,boyd2013chebyshev}.
We expand the solution $X_q(u,v)$ in a truncated series of $N$ Chebyshev polynomials in the variable $y=2u-1$, with coefficients depending on $v$. By requiring the equation to be satisfied in $N$ Gauss-Lobatto collocation points 
\begin{equation}
u_j=\frac{1}{2} \left(1-\cos\left(\frac{j \,\pi}{N-1}\right)\right) \qquad\qquad j=0,1,\ldots N-1\,,
\end{equation}
one is left with a system of $N$ (nonlinear) ordinary differential equations in the variable $v$, which we then solve using a standard Runge-Kutta algorithm. 
$N=30$ is used, and we have checked the final results  do not change for higher values of $N$.

In the following, the behavior of the chiral condensate as a function of time $v$ will be investigated and compared to the static result in two models, which assume two different functional forms for $z_h(v)$.
In Sec.~\ref{sec:power_law} a power-law behavior for $z_h(v)$ is assumed, while in Sec.~\ref{sec:hydro} we  use the metric discussed in \cite{Bellantuono:2015hxa,Bellantuono:2016tkh,Bellantuono:2017msk}, which is an analytic solution of Einstein equations in the large-time expansion and holographically reproduces the viscous hydrodynamics energy-momentum tensor of the gauge theory.
Results will be mostly shown in parametric plots of $\sigma(v)$ versus $T(v)$ to get a direct comparison with the static curve.
In all the considered cases the quark mass $m_q$ is assumed to be constant.

\subsection{Power-law behavior for $z_h(v)$}\label{sec:power_law}
For simplicity, we consider a toy model in which we assume that the metric functions are the same as in the static case, the only difference being that $z_h$ varies with time: $\Sigma(z,v)=1/z$, $A(z,v)=1/z^2 (1-z^4/z_h(v)^4)$ and $B(z,v)=0$. 
Although the system is not at thermodynamical equilibrium, we define an effective temperature $T(v)=1/(\pi z_h(v))$, coinciding with the temperature of a black-hole metric in the static case.
Two different profiles are used for $z_h(v)$:
\begin{equation}\label{eq:zhv}
    z_h(v)/z_h(0) = 1 \pm \Lambda_\ell \, (v/z_h(0))^{a} \qquad\qquad a\geqslant 1\,,
\end{equation}
corresponding to an effective temperature 
\begin{equation}\label{eq:Tv}
    T(v)/T(0) = (1 \pm   \Lambda_\ell \, (v/z_h(0))^{a})^{-1}\,,
\end{equation}
which is increasing ($-$) or decreasing ($+$) with time $v$.

\begin{figure}
    \begin{center}
        \begin{subfigure}[h]{0.49\textwidth}
            \centering
            \includegraphics[width=\textwidth]{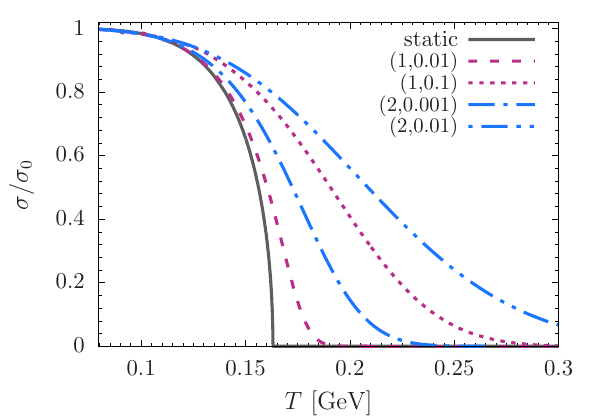}
            \label{fig:sigmazh4OS}
        \end{subfigure}
        \hfill
        \begin{subfigure}[h]{0.49\textwidth}
            \centering
            \includegraphics[width=\textwidth]{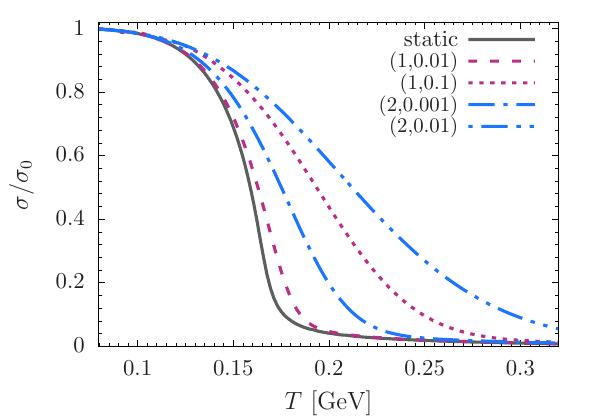}
            \label{fig:sigmazh4OSmq}
        \end{subfigure}
        \caption{Time-dependent chiral condensate as a function of the effective temperature $T(v)$ in Eq. \eqref{eq:Tv} (minus sign) for $m_q=0$ (left panel) and $m_q=3.22$ MeV (right panel), with the values of the parameters $(a,\Lambda_\ell)$ indicated in the legend. Time increases from left to right. The initial condition $X_q(u,0)$ is set at $z_h(0)=4$ GeV$^{-1}$, corresponding to $T(0)\sim 80$ MeV. $X_q(u,0)$ is the solution to Eq. \eqref{eq:eomstatic} at $z_h(0)$. The values of the chiral condensate  computed in the static case are shown by the gray plain curve. $\sigma_0$ is the value of $\sigma$ at $T=0$ at equilibrium.}
        \label{fig:sigmazh4OStot}
    \end{center}
\end{figure}

In the first case, we consider a decreasing $z_h(v)$, starting from $v=0$ and spanning the phase diagram from a state in which the temperature is initially small to higher values. 
We set $z_h(0)=4$ GeV$^{-1}$, corresponding to the low-temperature region of Fig. \ref{fig:sigmastatic} in which the chiral condensate is approximately constant. 
We assume the system is initially at equilibrium, so we require as initial condition that $X_q(u,0)$ is the solution computed at $z_h(0)$ in the static case. As boundary conditions we require that $X_q(0,v)=0$ and $\partial_u X_q(0,v)=m_q\, \gamma\, z_h(v)$. We also assume that the quark mass is constant.
At each fixed time $v$, $\sigma(v)$ is found, as in the static case, as the coefficient of the $z^3$ term in the low-$z$ expansion \eqref{eq:lowz} of the solution $X_q(u,v)$ with $u=z/z_h(v)$.
The obtained values of $\sigma(v)$ at the corresponding $T(v)$ are shown as parametric plots in Fig. \ref{fig:sigmazh4OStot} for $m_q=0$ (left panel) and $m_q=3.22$ MeV (right panel), and for $a=1$, $\Lambda_\ell=0.01, 0.1$, and $a=2$, $\Lambda_\ell=0.001, 0.01$, together with the values in the static case.
The transition to $\sigma=0$ is smooth for both $m_q=0$ and $m_q\neq 0$, as emphasized in Ref. \cite{Evans:2010xs}, and the curves in the two cases are similar. 
The curve with $\Lambda_\ell=0.01$ and $a=1$ is the closest to the static one, while they in general deviate more from the equilibrium curve as $a$ and $\Lambda_\ell$ increase. 
The effective temperature 
corresponding to the inflection point increases with $\Lambda_\ell$ and $a$, meaning that it increases as $T$ varies more rapidly with $v$.

\begin{figure}
    \begin{center}
        \begin{subfigure}[h]{0.49\textwidth}
            \centering
            \includegraphics[width=\textwidth]{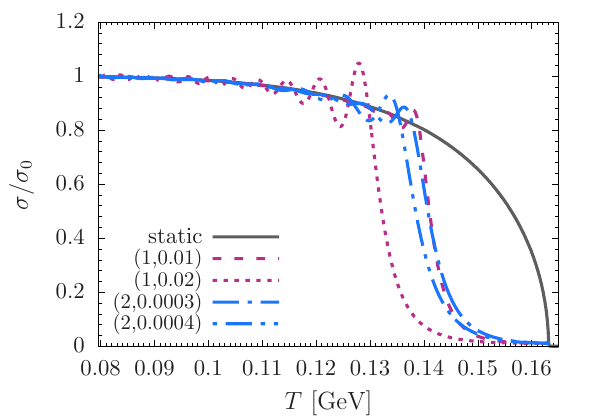}
            \label{fig:sigmadynamicLOW0}
        \end{subfigure}
        \hfill
        \begin{subfigure}[h]{0.49\textwidth}
            \centering
            \includegraphics[width=\textwidth]{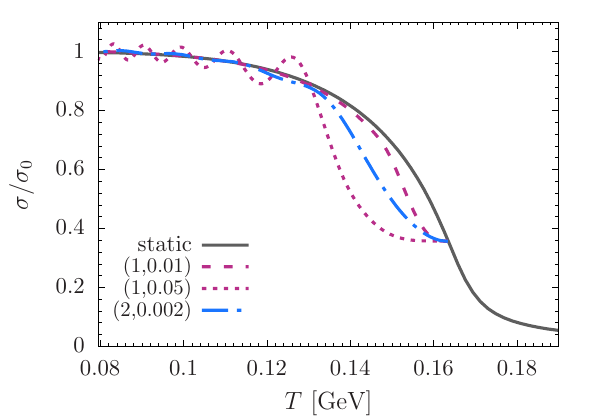}
            \label{fig:sigmadynamicLOWmq}
        \end{subfigure}
        \caption{Time-dependent chiral condensate as a function of $T(v)$ in Eq. \eqref{eq:Tv} (plus sign) for $m_q=0$ (left panel) and $m_q=3.22$ MeV (right panel), with the values of the parameters $(a,\Lambda_\ell)$ indicated in the legend.  Time increases from right to left. The initial condition $X_q(u,0)$ is set at $z_h(0)=1.949$ GeV$^{-1}$, corresponding to $T(0)\sim 163$ MeV. $X_q(u,0)$ is the solution to Eq. \eqref{eq:eomstatic} at $z_h(0)$. The values of the chiral condensate  computed in the static case are shown by the gray plain curve. $\sigma_0$ is the value of $\sigma$ at $T=0$ at equilibrium.}
        \label{fig:sigma1p95tot}
    \end{center}
\end{figure}

As a second example, we choose an increasing $z_h(v)$ according to Eq. \eqref{eq:zhv} with the plus sign, so the system is at high temperature at first, and then cools down as time goes by.
We set the initial condition at $z_h(0)=1.949$ GeV$^{-1}$, close to the critical temperature $T_c$ in the static case at $m_q=0$. 
We again assume that the system is initially at equilibrium, so the initial condition $X_q(u,0)$ is the solution computed at $z_h(0)$ in the static case.
If $\Lambda_\ell\to 0$ the static curve is reproduced by the dynamical evolution. 
Results for higher values of $\Lambda_\ell$ are shown in Fig. \ref{fig:sigma1p95tot}. 
Since we are considering the case in which temperature decreases with time, higher values of temperature correspond to initial times, while the temperature gets low at the end of the evolution.
At the beginning ($T\sim 160$ MeV) the system tends to stay close to the initial condition, so the chiral condensate remains small, and then it starts growing, running after the static curve. 
Eventually, the dynamical solution approaches the static one. 
At this stage oscillations can occur, which are damped as temperature further decreases. 
An analogous oscillatory behavior towards equilibrium has been observed in \cite{Cao:2022mep} when studying the evolution of the chiral condensate in a static background when the initial value of $\sigma$ is different from the on-shell one.
In \cite{Evans:2010xs} oscillatory solutions obtained with an excited initial condition, having an energy excess with respect to the on-shell one, are linked to bubble formation. In particular, a large homogeneous bubble forms in the mixed phase when there is an excess of energy due to a thermal fluctuation.
Therefore, it is also interesting to investigate what happens in a time-dependent background if the initial condition is different from the on-shell static one.
This analysis is described in the rest of the section and in the next one, using a different metric.

We now assume the scalar field is at some time perturbed from the equilibrium profile, getting the expression in Eq. \eqref{eq:lowz} with a given value of $\sigma$.

\begin{figure}
    \begin{center}
        \begin{subfigure}[h]{0.49\textwidth}
            \centering
            \includegraphics[width=\textwidth]{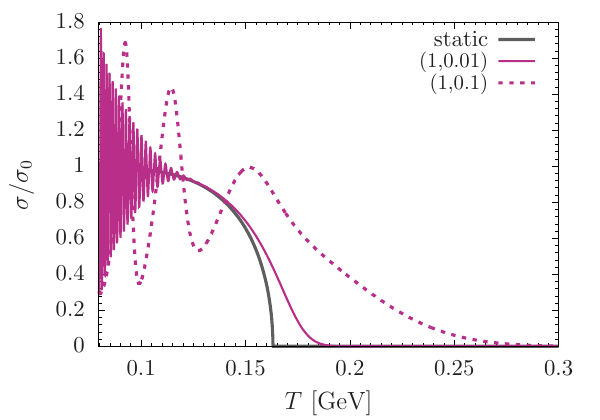}
            \label{fig:sigmazh4}
        \end{subfigure}
        \hfill
        \begin{subfigure}[h]{0.49\textwidth}
            \centering
            \includegraphics[width=\textwidth]{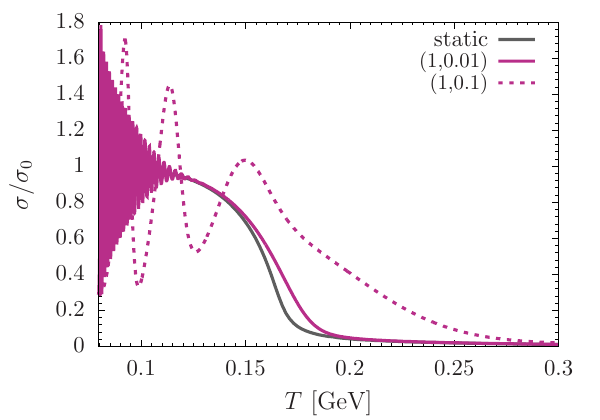}
            \label{fig:sigmazh4mq}
        \end{subfigure}
        \caption{Time-dependent chiral condensate as a function of $T(v)$ in Eq. \eqref{eq:Tv} (minus sign) for $m_q=0$ (left panel) and $m_q=3.22$ MeV (right panel), with the values of the parameters $(a,\Lambda_\ell)$ indicated in the legend.  Time increases from left to right. The initial condition $X_q(u,0)$ is set at $z_h(0)=4$ GeV$^{-1}$ and it is given by Eq.  \eqref{eq:lowz} with $\sigma=1/z_h(0)^3$. The values of the chiral condensate  computed in the static case are shown by the gray curve. $\sigma_0$ is the value of $\sigma$ at $T=0$ at equilibrium.}
        \label{fig:sigmazh4tot}
    \end{center}
\end{figure}

For an increasing temperature, the initial condition given by Eq. \eqref{eq:lowz} with $\sigma=1/z_h(0)^3$ is set at $z_h(0)=4$ GeV$^{-1}$.  The values of $\sigma$ obtained letting $T$ increase according to \eqref{eq:Tv} (minus sign) are shown in Fig. \ref{fig:sigmazh4tot} for $m_q=0$ (left panel) and $m_q=3.22$ MeV (right panel). 
Damped oscillations are observed in the beginning, which disappear when the system approaches the equibrium curve. In particular, for smaller $\Lambda_\ell$ the oscillations are faster and the system reaches equilibrium at lower temperatures.

\begin{figure}
    \begin{center}
    \includegraphics[width=9cm]{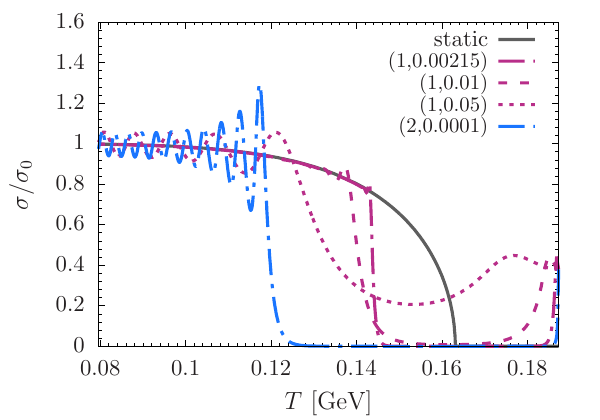}
    \end{center}
    \caption{Time-dependent chiral condensate as a function of $T(v)$ in Eq. \eqref{eq:Tv} (plus sign) for $m_q=0$, with the values of the parameters $(a,\Lambda_\ell)$ indicated in the legend. Time increases from right to left. The initial condition $X_q(u,0)=10^{-1} u^3$ is set at $z_h(0)=1.7$ GeV$^{-1}$, corresponding to $T(0)\sim 187$ MeV. The static result is shown by the plain gray curve. $\sigma_0$ is the value of $\sigma$ at $T=0$ at equilibrium.}
    \label{fig:sigmazhin1p7}
\end{figure}

\begin{figure}
    \begin{center}
    \includegraphics[width=9cm]{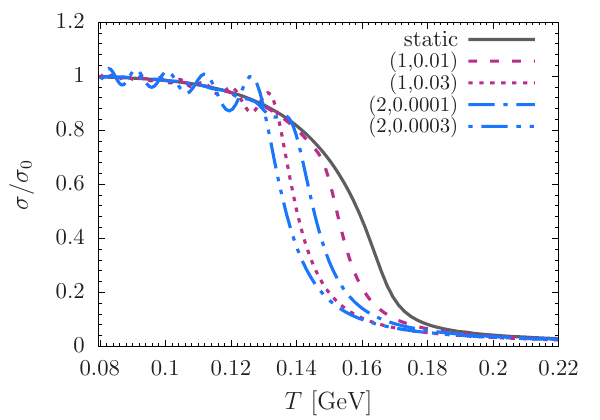}
    \end{center}
    \caption{Time-dependent chiral condensate as a function of $T(v)$ in Eq. \eqref{eq:Tv} (plus sign) for $m_q=3.22$ MeV, with the values of the parameters $(a,\Lambda_\ell)$ indicated in the legend. Time increases from right to left. The initial condition $X_q(u,0)=0$ is set at $T(0)=0.32$ GeV ($z_h(0)=1$ GeV$^{-1}$). The static result is shown by the gray plain curve. $\sigma_0$ is the value of $\sigma$ at $T=0$ at equilibrium.}
    \label{fig:sigmazhin1mq}
\end{figure}

In case of a decreasing temperature
and in the chiral limit, we start from $z_h(0)=1.7$ GeV$^{-1}$ (or $T(0)\sim 187$ MeV) and set the initial condition given by Eq. \eqref{eq:lowz} with $\sigma=0.1/z_h(0)^3$.  The values of $\sigma$ obtained letting $z_h$ increase according to \eqref{eq:zhv} (plus sign) are shown in Fig. \ref{fig:sigmazhin1p7}.
In general, in the chiral limit we find that if the initial condition is $X_q(u,0)=0$, the solution vanishes for any subsequent $v$.   
For $\Lambda_\ell=0.05$ and $a=1$, $\sigma$ is finite during the evolution.
In the other considered cases the system undergoes a transition (crossover) from very tiny values of $\sigma$ ($\sigma \sim 0$) to a phase in which chiral symmetry is broken, taking place at a temperature $T<T_c$. 
Such temperature  increases with decreasing $\Lambda_\ell$, and thermalization occurs at a higher $T$.
However, for even smaller $\Lambda_\ell$, in particular if $\Lambda_\ell<0.00215$ for $a=1$, at the beginning of the evolution, when $T>T_c$, the system thermalizes to $X_q=0$ and $\sigma=0$ (numerically $\sigma<10^{-13}$ GeV$^3$) and then they continue to be zero for subsequent times.
Oscillations are also observed.
In the $m_q\neq 0$ case, starting from the initial condition $X_q(u,0)=0$ at $z_h(0)=1$ GeV$^{-1}$, we observe that the system quickly reaches equilibrium or small oscillations are observed as long as the parameters $a$ and $\Lambda_\ell$ are small, \emph{i.e.} the variation of temperature is slow. This is shown in Fig. \ref{fig:sigmazhin1mq}. 
For smaller values of $\Lambda_\ell$ the dynamic curve approaches the static one at higher temperatures.
In the linear case $a=1$, if $\Lambda_\ell<0.001$ the process is quasi-static and the dynamic curve coincides with the static one.
The effective temperature corresponding to the inflection point is smaller than the one in the static case, and it decreases with increasing $\Lambda_\ell$ and $a$.


\begin{figure}
    \begin{center}
    \includegraphics[width=9cm]{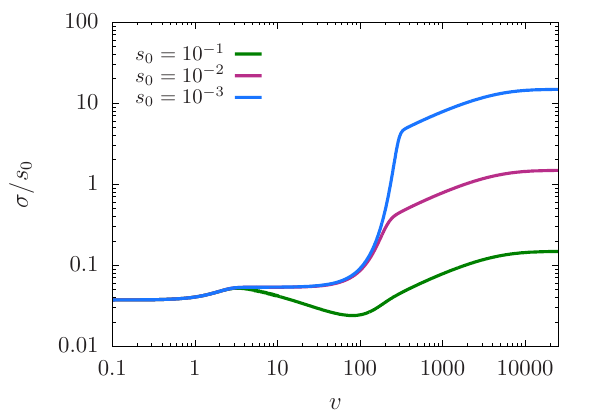}
    \end{center}
    \caption{Time-dependent chiral condensate for $m_q=0$ in a blackground with decreasing temperature according to \eqref{eq:Tv} (plus sign) with $a=1$, $\Lambda_\ell=10^{-4}$ and $z_h(0)=1.949$ GeV$^{-1}$, corresponding to $T(0)\sim T_c$. The initial condition is $X_q(u,0)=s_0 u^3$, with the values of $s_0$ indicated in the legend.}
    \label{fig:prethermLm4}
\end{figure}

\begin{figure}
    \begin{center}
    \includegraphics[width=9cm]{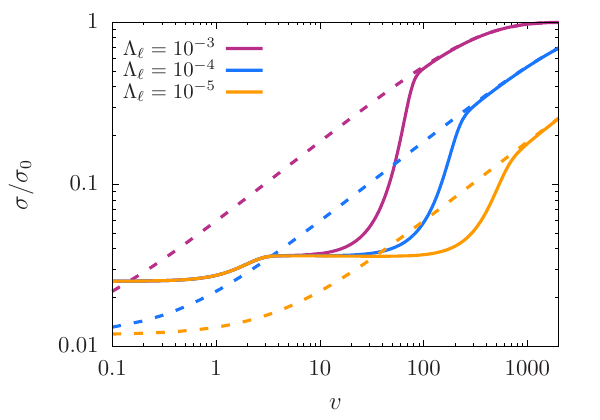}
    \end{center}
    \caption{Time-dependent chiral condensate  for $m_q=0$ in a blackground with decreasing temperature according to \eqref{eq:Tv} (plus sign) with $a=1$, $z_h(0)=1.949$ GeV$^{-1}$ and the values of $\Lambda_\ell$ indicated in the legend. The initial condition is $X_q(u)=s_0 u^3$, with $s_0=10^{-2}$. The static results corresponding to each value of $\Lambda_\ell$ are shown by the dashed curves. $\sigma_0$ is the value of $\sigma$ at $T=0$ at equilibrium.}
    \label{fig:prethermsm2}
\end{figure}

In out-of-equilibrium studies a prethermalization phase has been found, in which the ratio of pressure over energy density is almost constant although the system is still not equilibrated  \cite{Berges:2004ce}. This has been also observed when investigating collision of shock waves using gauge/gravity duality producing a far-from-equilibrium strongly-coupled plasma which relaxes towards a hydrodynamic regime, see \emph{e.g.} \cite{Chesler:2009cy,Bellantuono:2015hxa}. 
In general a prethermalization phase is a transient quasistationary state different from the true thermal equilibrium \cite{Berges:2004ce} while after a sufficiently long time the system thermalizes.
In this phase the memory of the initial state is partially kept, while it is completely lost in the thermal phase \cite{Saha:2024eto}. 
Prethermalization usually occurs when there are relevant separated  time scales and relaxation proceeds in several steps \cite{Mori:2017qhg}.
Prethermalization and thermalization phases have been studied in AdS/QCD in \cite{Cao:2022mep} in the case of a system at fixed temperature quenched to a perturbed initial state, and the interplay between order parameter and the corresponding Goldstone modes has been investigated in \cite{Zheng:2024rzl}.
The system considered here is in contact with an evolving background, in which temperature changes with time. In the chiral limit, if the initial temperature is very close to the critical one $T_c$ and if $z_h$ slowly changes with time, the initial behavior of the system is akin to a prethermalization phase. 
This is shown in Figs. \ref{fig:prethermLm4} and \ref{fig:prethermsm2}. In Fig. \ref{fig:prethermLm4} the time evolution of the chiral condensate starts from $z_h(0)=1.949$ GeV$^{-1}$, corresponding to $T(0)\sim T_c$, with $X_q(u,0)=s_0 u^3$ and $s_0=10^{-1}, 10^{-2}, 10^{-3}$. Temperature and black-hole horizon change with time according to Eqs. \eqref{eq:Tv}-\eqref{eq:zhv}, respectively, with $a=1$ and $\Lambda_\ell=10^{-4}$. A prethermalization stage, in which the system is almost stationary and trapped in a state different from the equilibrium one, can be observed from $v \sim 4$ GeV$^{-1}$, and it lasts more as $s_0$ gets closer to the equilibrium value.
As shown in Ref. \cite{Cao:2022mep} during the prethermalization stage the ratio $\sigma/s_0$ is almost constant and does not depend on $s_0$.
In Fig. \ref{fig:prethermsm2} the dependence on $\Lambda_\ell$ is investigated, showing that the system remains in the prethermalization phase longer for smaller values of $\Lambda_\ell$. 
Hence, the duration of the prethermalization phase depends on how fast the temperature changes with time and how much the initial condition deviates from the equilibrium one.
Equilibrium, represented by the dashed lines, is eventually reached.
For $m_q\neq 0$ we find no evidence of prethermalization.


\subsection{Hydrodynamics-like behavior for $z_h(v)$}\label{sec:hydro}
We now consider a different time-dependent metric, which has been obtained in Refs. \cite{Bellantuono:2015hxa,Bellantuono:2016tkh,Bellantuono:2017msk} to describe a viscous hydrodynamic system in a holographic model.
It has been found (see, e.g., \cite{Chesler:2008hg}) that if an external source perturbes (quenches) a boost-invariant boundary metric within a definite time interval, a horizon appears in the bulk. 
The quench, represented by the profile $\gamma(v)$, modifies the line element of the $4d$ boundary:
\begin{equation}
    ds^2_4=-dv^2+e^{\gamma(v)} dx_\perp^2+v^2 e^{-2\gamma(v)} dy^2 \,\, , \label{metric4D}
\end{equation}
leaving the spatial three-volume unchanged and preserving the translational and $O(2)$ symmetries in the transverse plane.
The 5$d$ spacetime on which the gravity dual is defined, having the metric \eqref{metric4D} as boundary, is described by the line element in Eq. \eqref{metric5D}.
It has also been observed that, after a very short time from the end of the quench,  the system thermalizes reaching a behavior which is well described by viscous hydrodynamics.
Indeed, the energy-momentum tensor of the gauge theory, computed from the  $5d$ metric using the holographic renormalization procedure developed in \cite{deHaro:2000vlm}, approaches soon after the end of the quench the energy-momentum tensor obtained in viscous hydrodynamics through a Bjorken expansion \cite{Heller:2012je}:
\begin{equation}
    T^\mu_\nu=\frac{N_c^2}{2\pi^2}\mbox{diag}(-\epsilon,p_\perp,p_\perp,p_{||})
\end{equation}
with
\begin{eqnarray}
\epsilon(v)&=& \frac{3 \pi^4 \Lambda^4}{4 (\Lambda v)^{4/3} }\left[ 1-\frac{2c_1}{ (\Lambda v)^{2/3}}+\frac{c_2}{ (\Lambda v)^{4/3}} + {\cal O}\left( \frac{1}{(\Lambda v)^2} \right )\right] \label{hydroeps} \\
p_\parallel (v)&=&\frac{ \pi^4 \Lambda^4}{ 4(\Lambda v)^{4/3} } \left[ 1-\frac{6c_1}{ (\Lambda v)^{2/3}}+\frac{5c_2}{ (\Lambda v)^{4/3}} + {\cal O}\left( \frac{1}{(\Lambda v)^2} \right )\right] \label{hydroppar} \\
p_\perp (v) &=& \frac{ \pi^4 \Lambda^4}{ 4(\Lambda v)^{4/3} } \left[ 1-\frac{c_2}{ (\Lambda v)^{4/3}} + {\cal O}\left( \frac{1}{(\Lambda v)^2} \right ) \right] \,\,\, ,\label{hydropperp}
\end{eqnarray}
where $c_1=\displaystyle{\frac{1}{3 \pi}}$, $c_2=\displaystyle{\frac{1+2 \log{2}}{18 \pi^2}}$, and $\Lambda$ is a parameter. 
$\epsilon$ is the energy density, $p_{||}$ is the pressure in the longitudinal direction along which the plasma expands, $p_\perp$ is the pressure along one of the two transverse directions.
Such hydrodynamic model is able to mimic the behavior of the QGP in a relativistic heavy-ion collision.
The metric that produces such an energy-momentum tensor, solution of Einstein equations in the late time expansion,  is \eqref{metric5D} with \cite{Bellantuono:2016tkh}:
\begin{eqnarray} 
A(z,v) &=& \frac{1}{z^2} \left(1 - \frac{4 z^4}{3} \epsilon(v) \right)  \\
\Sigma(z,v) &=&\frac{1}{z} \left(v+z\right)^{1/3} \\
B(z,v) &=& \frac{z^4}{3} (p_\perp(v) -p_{||}(v))-\frac{2}{3}\log(v+z)\,. \label{eq:metrichydro}
\end{eqnarray}
The effective temperature, defined as  
\begin{equation}
    \epsilon(v)=\frac{3}{4}\pi^4 T(v)^4\,,
\end{equation}
is given by:
\begin{eqnarray}
    T(v)&=&\frac{\Lambda}{(\Lambda v)^{1/3}} \Bigg[ 1-\frac{1}{6 \pi (\Lambda v)^{2/3}}+\frac{-1+\log 2}{36 \pi^2 (\Lambda v)^{4/3} } + {\cal O}\left( \frac{1}{(\Lambda v)^{2}} \right )\Bigg]\,,  \label{Teff1}
\end{eqnarray}
and the position of the horizon (\emph{i.e.} the value of $z$ at which $A(z,v)$ vanishes) by:
\begin{eqnarray}
    z_{h}(v)&=&\frac{(\Lambda v)^{1/3}}{\pi \Lambda} \Bigg[ 1+\frac{c_1}{2 (\Lambda v)^{2/3}}+\frac{5c_1^2-2 c_2}{8 (\Lambda v)^{4/3} } + {\cal O}\left( \frac{1}{(\Lambda v)^{2}} \right )\Bigg]\,.  \label{zheff}
\end{eqnarray}
Four models have been considered in \cite{Bellantuono:2015hxa}, each characterised by a different form of the quench. The parameter $\Lambda$ takes the following values: $\Lambda=2.25$ GeV in model $\mathcal{A}_1$, $\Lambda=1.73$ GeV in model $\mathcal{A}_2$, $\Lambda=1.12$ GeV in model $\cal{B}$, and $\Lambda=1.59$ GeV in model $\cal{C}$. 
Notice that in \cite{Bellantuono:2015hxa} no energy scale was set and no conformal symmetry breaking scheme was introduced, so the values of $\Lambda$ were dimensionless. 
In this work we have introduced an energy scale through a static dilaton, and all quantities get a scale dimension.
In the following we shall study the scalar field in the background described by Eqs. \eqref{hydroeps}-\eqref{eq:metrichydro}, with the above four values of $\Lambda$.

\begin{figure}
    \begin{center}
    \includegraphics[width=9cm]{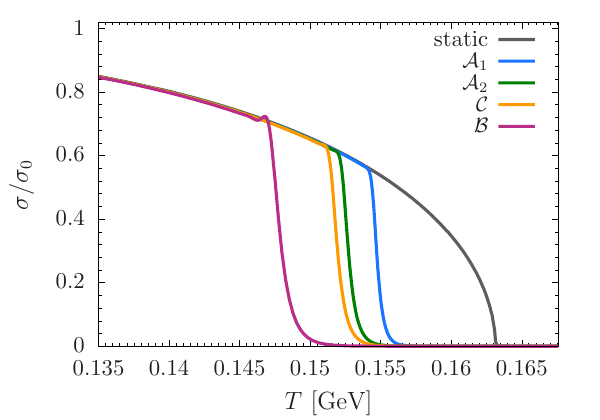}
    \end{center}
    \caption{Time-dependent chiral condensate as a function of the effective temperature $T(v)$ in \eqref{Teff1} in the hydrodynamic model described by Eqs.  \eqref{hydroeps}-\eqref{eq:metrichydro} with $m_q=0$. The initial condition $X_q(u,v_0)=10^{-5} u^3$ is set at $z_h(v_0)=1.9$ GeV$^{-1}$, corresponding to $T(v_0)\sim 168$ MeV. Time increases from right to left. The static result is shown by the gray curve.}
    \label{fig:sigmaHydrozhin1p9}
\end{figure}

In the chiral limit, we have set the initial condition $X_q(u,v_0)=10^{-5} u^3$ at initial time $v_0$ corresponding to $z_h(v_0)=1.9$ GeV$^{-1}$ (or $T(v_0)\sim 168$ MeV). Temperature decreases with increasing time $v$. Solving the partial differential equation and fitting the solution to Eq. \eqref{eq:lowz}, we have obtained the values of $\sigma$ shown in Fig. \ref{fig:sigmaHydrozhin1p9}. 
In general, $z_h(v)=z_0$ has two real solutions and, since we are considering the large-time limit, we take the largest value of $v$. 
As found in the previous section, we again find that in all these models, for $m_q=0$, if $X_q(u,v_0)=0$ is set for any $v_0$ or if a finite $X_q$ is set as  initial condition at a temperature sufficiently higher than the critical one, the solution vanishes for all times. 
Therefore, a transition to a chirally broken phase is triggered by fluctuations of the scalar field $X_q\neq 0$ for temperatures close to the critical one or lower. 
Fig.  \ref{fig:sigmaHydrozhin1p9} shows that, if a fluctuation of the scalar field occurs at $T\gtrsim T_c$, the system takes some time to reach equilibrium and the transition to a chirally broken phase starts at $T<T_c$. Once $\sigma$ gets its equilibrium value, it continues to follow the equilibrium curve as the temperature decreases. 
We have checked that this behavior is also realised when the initial fluctuation is larger than the considered one. 

\begin{figure}
    \begin{center}
    \includegraphics[width=9cm]{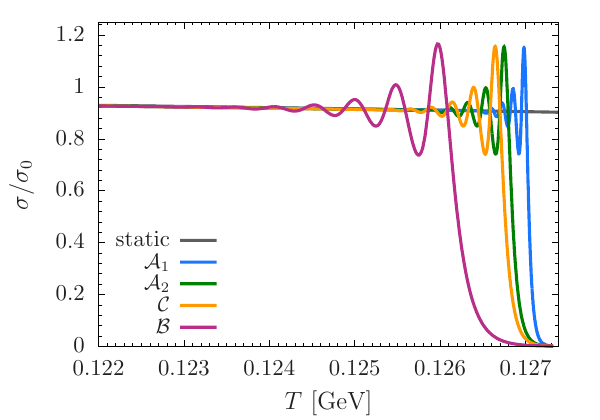}
    \end{center}
    \caption{Time-dependent chiral condensate as a function of the effective temperature $T(v)$ \eqref{Teff1}  in the hydrodynamic model described by Eqs.  \eqref{hydroeps}-\eqref{eq:metrichydro} with $m_q=0$. The initial condition $X_q(u,v_0)=0.001 u^3$ is set at $z_h(v_0)=2.5$ GeV$^{-1}$, corresponding to $T(v_0)\sim 127.3$ MeV. Time increases from right to left. The static result is shown by the gray curve.}
    \label{fig:oscillation}
\end{figure}

If a fluctuation of $X_q$ occurs at a much lower temperature, short oscillations are observed before equilibrium is reached. 
This is shown in Fig. \ref{fig:oscillation}, obtained setting the initial condition $X_q(u,v_0)=0.001 u^3$ at $z_h(v_0)=2.5$ GeV$^{-1}$, corresponding to an intial temperature $T(v_0)=127.3$ MeV.
The height of the first oscillation is the same in the four models, while the amplitude is inversely proportional to $\Lambda$, and model $\mathcal{B}$, which has the smallest $\Lambda$, reaches equilibrium at lower temperatures. 
Notice that, according to Eq. \eqref{zheff},  $z_h(v)$  varies more slowly with time when  $\Lambda$ is larger.  

\begin{figure}
    \begin{center}
    \includegraphics[width=9cm]{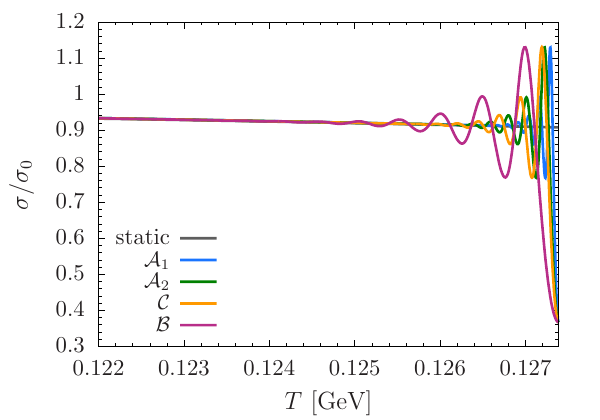}
    \end{center}
    \caption{Time-dependent chiral condensate as a function of $T(v)$ in the hydrodynamic model described by Eqs.  \eqref{hydroeps}-\eqref{eq:metrichydro} with $m_q=3.22$ MeV. The initial condition $X_q(u,v_0)$ is given by Eq. \eqref{eq:lowz} with $\sigma=0.02$ GeV$^3$ and is set at $z_h(v_0)=2.5$ GeV$^{-1}$, corresponding to $T(v_0)=127.3$ MeV. Time increases from right to left. The static result is shown by the gray curve.}
    \label{fig:sigmaHzhin2p5mq}
\end{figure}

When $m_q=3.22$ MeV, we again find that any initial condition for the scalar field leads to a chirally broken phase at low temperature, and $\sigma(T)$ rapidly approaches the static value. In particular, if the initial state is $X_q(u,v_0)=0$ at $z_h(v_0)=1$ GeV$^{-1}$, the system behaves as at equilibrium during the whole evolution in all the models. 
This also occurs if $X_q(u,v_0)\neq 0$ since thermalization turns out to be very rapid.
For higher $z_h(v_0)$, if the initial $\sigma(v_0)$ is different from the expected value in the static limit, a short oscillation can form, after which the static values are reproduced. 
In this respect, we have investigated what happens if at some point during the crossover the value of $\sigma$ is modified by some external effect. 
To this aim, we have solved the equation of motion for $X_q$ setting an initial condition at $z_h(v_0)=2.5$  GeV$^{-1}$ such that $X_q$ is equal to the asymptotic expression in Eq. \eqref{eq:lowz} with $\sigma=0.02$ GeV$^{-3}$, a value different from the equilibrium one. Results reported in  Fig. \ref{fig:sigmaHzhin2p5mq} show that, after a very short oscillation, the system returns to equilibrium.

\begin{figure}
    \begin{center}
    \includegraphics[width=9cm]{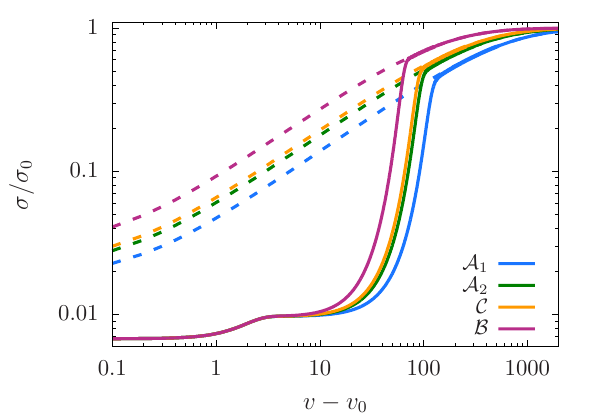}
    \end{center}
    \caption{Time-dependent chiral condensate as a function of $v-v_0$ in the hydrodynamic model described by Eqs.  \eqref{hydroeps}-\eqref{eq:metrichydro} with $m_q=0$. The initial condition $X_q(u,v_0)=10^{-4} u^3$ is set close to $T_c$ at $z_h(v_0)=1.949$ GeV$^{-1}$. The static results are shown by the dashed curves.}
    \label{fig:prethermH}
\end{figure}

It is possible to observe prethermalization in this system in the chiral limit by choosing an initial temperature close to the critical one $T_c$. In Fig. \ref{fig:prethermH} $\sigma$ is plotted as a function of $v-v_0$, where $v_0$ is the initial time, such that $z_h(v_0)=1.949$ GeV$^{-1}$. For all the models, the initial condition is $X_q(u,v_0)=10^{-4} u^3$.
Model $\mathcal{B}$, having the smallest $\Lambda$, has the shortest prethermalization stage, in agreement with the results found in the previous section.

\section{Conclusions}\label{sec:conclusions}
We have studied the chiral condensate at finite temperature in AdS/QCD in a time-dependent background, with 
conformal invariance broken by a static dilaton, as in the soft-wall model. 
A quartic potential term  and the anomalous dimension of the scalar field $X_q$ allow us to study chiral symmetry breaking for massless quarks.
We have considered two kinds of initial conditions for the evolution equation: the equilibrium on-shell solution of $X_q$ and  a perturbed profile, where the initial chiral condensate is abruptly shifted from its equilibrium value.  

A toy model has been analysed first, assuming a general power-law time dependence for the position of the black-hole horizon $z_h(v)$, describing an increasing or decreasing temperature.
The parametric curve representing the evolution of the chiral condesate as a function of temperature remains smooth for both $m_q=0$ and $m_q\neq 0$.
Its deviation from the static case increases as the rate of temperature change over time becomes more pronounced. 
In the chiral limit, if the chiral condensate initially vanishes, the transition to $\sigma\neq 0$ does not occur. 
For $m_q\neq 0$, if $z_h$ varies slowly with time, $X_q$ evolves through quasistatic transformations (\emph{i.e.} adiabatically), so $\sigma$ gets almost the same values as in the static case, and this is roughly independent of the initial condition, since thermalization is very quick. 
In the chiral limit, the dependence on the initial condition, for a given $z_h(v)$, is subtler.
If the system is perturbed at $T>T_c$ with a small initial quench, where $T_c$ is the critical temperature characterizing the second-order transition in the static case, the system soon thermalizes to $\sigma=0$, and then it needs some time to reach the static behavior at low temperatures, in which $\sigma\neq 0$.
The smaller the initial $\sigma$, the higher the thermalization time for a given $z_h(v)$.
If instead the system is perturbed at $T<T_c$, thermalization time decreases when the initial condition is closer to the on-shell one.
We have also observed that at low temperatures the chiral condensate  oscillates around the equilibrium value when the solution for $X_q$ approaches the static one with an energy excess.

Then, we have considered a specific metric describing a cooling system in which the energy-momentum tensor  is the one found in viscous hydrodynamics. Also in this case, $z_h(v)$ follows a power-law behavior, but with a noninteger exponent and with corrections containing viscous effects.  
In the hydrodynamic setup the evolution of the medium is slow enough, and the chiral condensate computed in a dynamical model, solving partial differential equations, is almost equal to the one computed assuming the system is at equilibrium. Oscillations are observed if the initial condition for the scalar field  is different from the equilibrium value at $T<T_c$, which however soon disappear, and then the system behaves as at equilibrium.

A prethermalization stage, \emph{i.e.} a transient quasistationary state different from the true thermal equilibrium, is found in the chiral limit by choosing an initial temperature close to the critical temperature determined in the static case.

It would be interesting to study a system with dynamical dilaton, coupled to the metric and obtained, together with the metric functions, by solving time-dependent Einstein equations, and check if the general behavior described here is confirmed. 
We leave this for future work.

\section*{Acknowledgments}
We thank Loredana Bellantuono, Pietro Colangelo and Fulvia De Fazio for fruitful discussions and earlier collaboration on some of the models used in the present work.
The work is carried out within the INFN project (Iniziativa Specifica) SPIF.
It has been partly funded by the European Union - Next Generation EU through the research grant number P2022Z4P4B ``SOPHYA - Sustainable Optimised PHYsics Algorithms: fundamental physics to build an advanced society'' under the program PRIN 2022 PNRR of the Italian Ministero dell'Universit\`a e Ricerca (MUR).

\bibliographystyle{JHEP}
\bibliography{text.bib}

\end{document}